# Composite Cell from Nematic and Cholesteric Liquid Crystals as an Rotator of Electrically Tunable Polarization Plane

A.H. Gevorgyan [1], K.B. Oganesyan [2,3*], E.A. Ayryan [3], M. Hnatic [4,5,6], J.Busa [7], E. Aliyev [8], A.M. Khvedelidze [3,9,10], Yu.V. Rostovtsev [11], G. Kurizki [12]

[1] Yerevan State University, Yerevan, Armenia
[2] Alikhanian National Lab, Yerevan Physics Institute, Yerevan, Armenia
[3] LIT, Joint Institute for Nuclear Research, Dubna, Russia
[4] Faculty of Sciences, P. J. Safarik University, Park Angelinum 9, 041 54 Kosice, Slovakia
[5] Institute of Experimental Physics SAS, Watsonova 47, 040 01 Kosice, Slovakia ˇ
[6] BLTP, Joint Institute for Nuclear Research, Dubna, Russia
[7] Department of Mathematics and Theoretical Informatics, FEE&I, Technical University, Košice, Slovakia
[8] The Institute of Combustion Problems, Almaty, Kazakhstan
[9] A.Razmadze Mathematical Institute, Iv.Javakhishvili, Tbilisi State University, Tbilisi, Georgia
[10] Institute of Quantum Physics and Engineering Technologies, Georgian Technical University, Tbilisi, Georgia
[11] University of North Texas, Denton, TX, USA
[12] Weizmann Institute of Science, Rehovot, Israel

[*] bsk@yerphi.am

**Abstract**. A liquid crystal optical device made of an optically anisotropic heterostructure is considered. The device consists of a cholesteric liquid crystal (CLC) layer sandwiched by two phase-shifting anisotropic layers of a nematic liquid crystal (NLC). In this structure each of the NLC layers is a quarterwave plate. The problem is solved both by Ambartsumian's method of layer addition and Muller's matrix method. The peculiarities of reflection spectra, eigen polarizations, rotation of polarization plane and polarization ellipticity are studied. It is shown that this device can work as a light modulator or a system for obtaining linearly polarized light with electrically tunable rotation of the polarization plane (which is especially important for optical communication), as well as a device for obtaining the linearly polarized light from a non-polarized one.

## 1. INTRODUCTION

Multilayer liquid-crystal cells are the most known structural elements for development of modern technologies on electro-magneto-acousto-optical systems. These cells have a wide application as tunable polarization plane rotators [1–6] as well as dynamical phase shifters, achromatic miniaturized liquidcrystal devices for displays [7, 8], controlled filters [9–11], mirrorless dye-lasers [12–15], optical diodes [16, 17], etc. The main features of these devices are their easy controllability, low losses, and small sizes. In [17–21] novel optical heterojunctions of anisotropic structures, consisting of NLC layer sandwiched by two CLC layers, have been investigated. In these works an anisotropic layer, which is a half-wave plate, was considered. This is similar to two stage undulators [22-73].

In the present paper we study an NLC–CLC–NLC heterojunction, in which the anisotropic layer is a quarter-wave plate, and show that such a system also possesses unique properties.

## 2. RESULTS AND DISCUSSION

Consider the reflection and transmission of light through an NLC–CLC–NLC system (Fig. 1). The transmission of a plane polarized wave through such a system can be analyzed by the modified Ambartsumian's layer addition method [19,74]. Let a wave with a complex amplitude $\mathbf{E}_i$ fall normally to this system. Denoting the complex amplitudes of the reflected and transmitted fields by $\mathbf{E}_r$ and $\mathbf{E}_t$, expanding them in basis $p$- and $s$-polarizations, $\mathbf{E}_{i,r,t} = E^p_{i,r,t}\mathbf{n}_p + E^s_{i,r,t}\mathbf{n}_s = \begin{pmatrix} E^p_{i,r,t} \\ E^s_{i,r,t} \end{pmatrix}$, ($\mathbf{n}_p$ and $\mathbf{n}_s$ are the unit vectors of the $p$- and $s$-polarizations), we can represent the solution of problem as

$$\mathbf{E}_r = \hat{R}\mathbf{E}_i, \quad \mathbf{E}_t = \hat{T}\mathbf{E}_i \qquad (1)$$

where $\hat{R}$ and $\hat{T}$ are 2×2 matrices of reflection and transmission of this system. According to [19,74], if one has a system consisting of two adjacent "from left to right" layers A and B, then the matrices of reflection and transmission of the system A+B are defined in terms of analogous matrices of constituent layers by the following matrix equations:

$$\hat{R}_{A+B} = \hat{R}_A + \tilde{\hat{T}}_A R_B \left(\hat{I} - \tilde{\hat{R}}_A \hat{R}_B\right)^{-1} \hat{T}_A, \quad \hat{T}_{A+B} = \hat{T}_B \left(\hat{I} - \tilde{\hat{R}}_A \hat{R}_B\right)^{-1} \hat{T}_A, \qquad (2)$$

where $\hat{I}$ is the unit matrix, tilda denotes the corresponding matrices of reflection and transmission in the case of the opposite direction of propagation. To derive the reflection and transmission matrices of the NLC–CLC–NLC system, one can, using formula (2), first join a

CLC layer to the NLC layer from its left side and then join the second NLC layer to the obtained system again from the left side.

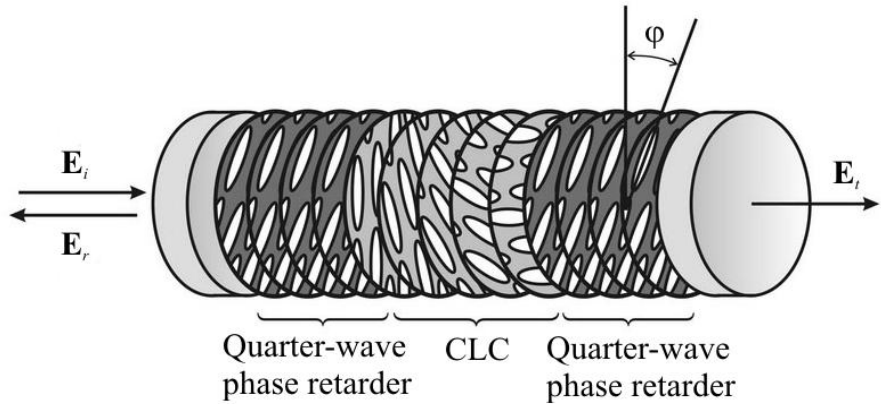

**Fig. 1.** Scheme of an optical heterostructure consisting of a CLC layer sandwiched by two NLC layers.

Now we proceed to consideration of obtained results. Figure 2 shows the reflection spectra at the incidence of light with the orthogonal linear and circular polarizations, with eigen polarizations (EPs), as well as the spectra of azimuth and ellipticity of the first and second EPs. The EPs are two polarizations of the incident wave which do not change in transmission of light through the system [74,75]. EPs and eigenvalues (amplitude coefficients of transmission at the incidence of light with EPs on the system) give a great body of information on the features of interaction of light with the system, therefore the calculation of EPs and eigenvalues for each given optical system is of great importance in optics. From the definition of EPs it follows that they should be connected with the polarization of the internal waves (eigenmodes) excited in the medium. In the general case, there are certain distributions: only two EPs exist, whereas the number of eigenmodes may be more than two and the polarizations of all these modes may differ from one another (for instance, in nonreciprocal media). Note that in EPs the influence of dielectric boundaries is taken into account. In particular, as known, at the normal incidence of light EPs of CLCs and gyrotropic media practically coincide with the orthogonal circular polarizations, whereas for nongyrotropic media they coincide with the orthogonal linear polarizations. It follows from the above that the investigation of features of EPs is especially important in the case of inhomogeneous media for which the exact solution of the problem is unknown.

As seen in Fig. 2, in contrast to the case of a separate CLC layer (which has a diffractive reflection selective with respect to the circular polarizations, i.e. in the photonic forbidden gap (PFB) the light with a circular polarization is completely reflected, while the light with the opposite circular polarization completely transmits), this system manifests the selectivity with respect to the linear polarizations: the light with a linear $x$-polarization undergoes a total reflection in the PFB, while the light with a linear $y$-polarization completely transmits. These properties of the system considered can be explained by the following circumstances. In propagation of light through an anisotropic crystal an additional phase difference appears due to the anisotropy. Because of this, if the crystal is a quarter-wave plate, then at the incidence of the linearly polarized light, the polarization plane of which makes an angle $\phi = \pi/4$ ( $\phi = -\pi/4$ ) with the optical axis of the plate, the transmitted light has a right (left) circular polarization.
In addition, this system has one more unique property, namely, its EPs are not orthogonal (or quasiorthogonal) linear, not circular or elliptical polarizations. In this case both EPs in the PFB have practically the linear (along the $x$-axis) polarization (Fig. 2c,d), and almost total reflection of light with both EPs in the PFB (Fig. 2b) is explained just by this fact.

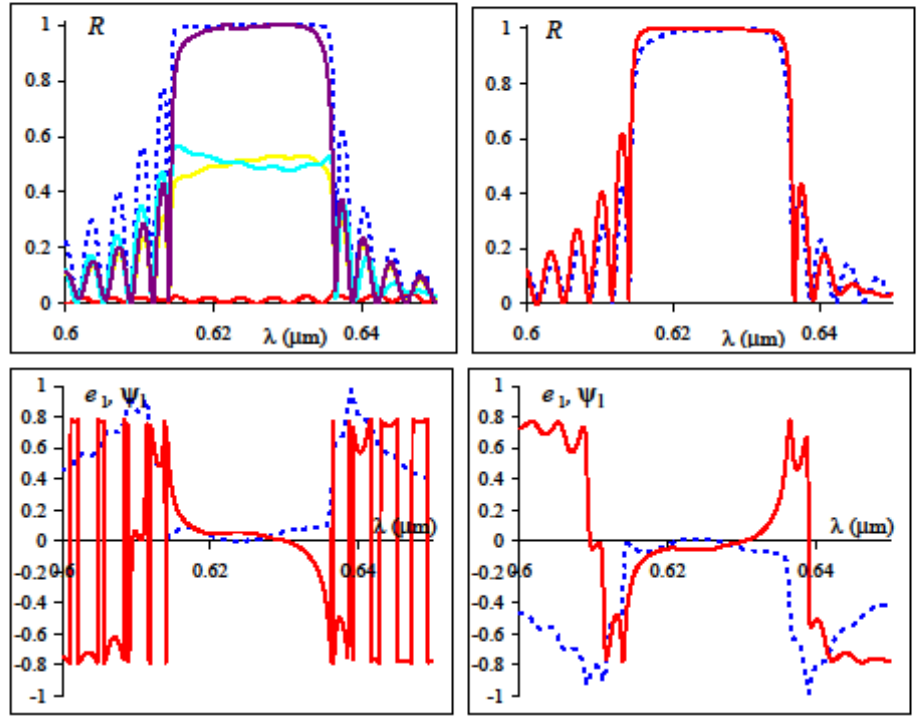

**Fig. 2.** Reflection spectra at the incidence on the NLC–CLC–NLC system of light with (a) orthogonal linear and circular polarizations, (b) with EPs, and spectra of (*1*) the azimuth and (*2*) ellipticity of (c) the first and (d) second EPs. The incident light has (a) the linear polarization along the axes (*1*) x and (*2*) y, and the circular (*3*) right and (*4*) left polarizations; (b) eigen polarizations. The CLC helix is right. The parameters of the CLC layer are the following: the principal values of the local tensor of dielectric permittivity 1 $\varepsilon$ = 2.29, 2 $\varepsilon$ = 2.143, the helix pitch $\sigma$ = 0.42 μm, the width $d = 70\sigma$. The parameters of the NLC layer are the following: the refractive indices 1.746, $e$ $n = 0$ $n$ =1.522, the width $d$ = 0.65 μm. The angle $\phi$ = 45°.

Consider now the influence of the angle $\phi$ (angle between the optical axis of the anisotropic crystal and the *x*-axis of laboratory system) on the reflection. We present the three-dimensional plots of the dependences of reflection on the wavelength and angle $\phi$ for the NLC–CLC–NLC system in Fig. 3. The incident light has the linear polarization along the *x*- and *y*-axis, as well as the right and left circular polarizations. As seen from the polarization plots, in the case of linear polarization of light the variation of $\phi$ essentially influences the light reflection in the PFB. In the case of light with the polarization linear along the *x*(*y*)-axis the reflection coefficient $R$ is changed from 0 to 1 (from 1 to 0) with variation of the angle $\phi$ from $-\pi/4$ to $\pi/4$. This means that such a system can operate as an ideal modulator with a modulation depth equal to 1.

Since one can control the orientation of axes of NLC molecules (and, hence, the orientation of the optical axis of the NLC layer) by an external electrical filed, then the NLC–CLC–NLC system may be used as an electrically controlled modulator. Variation of the angle $\phi$ does not influence the reflection in the case when the light with a circular (both left and right) polarization is incident on the system (see Fig. 3c,d).

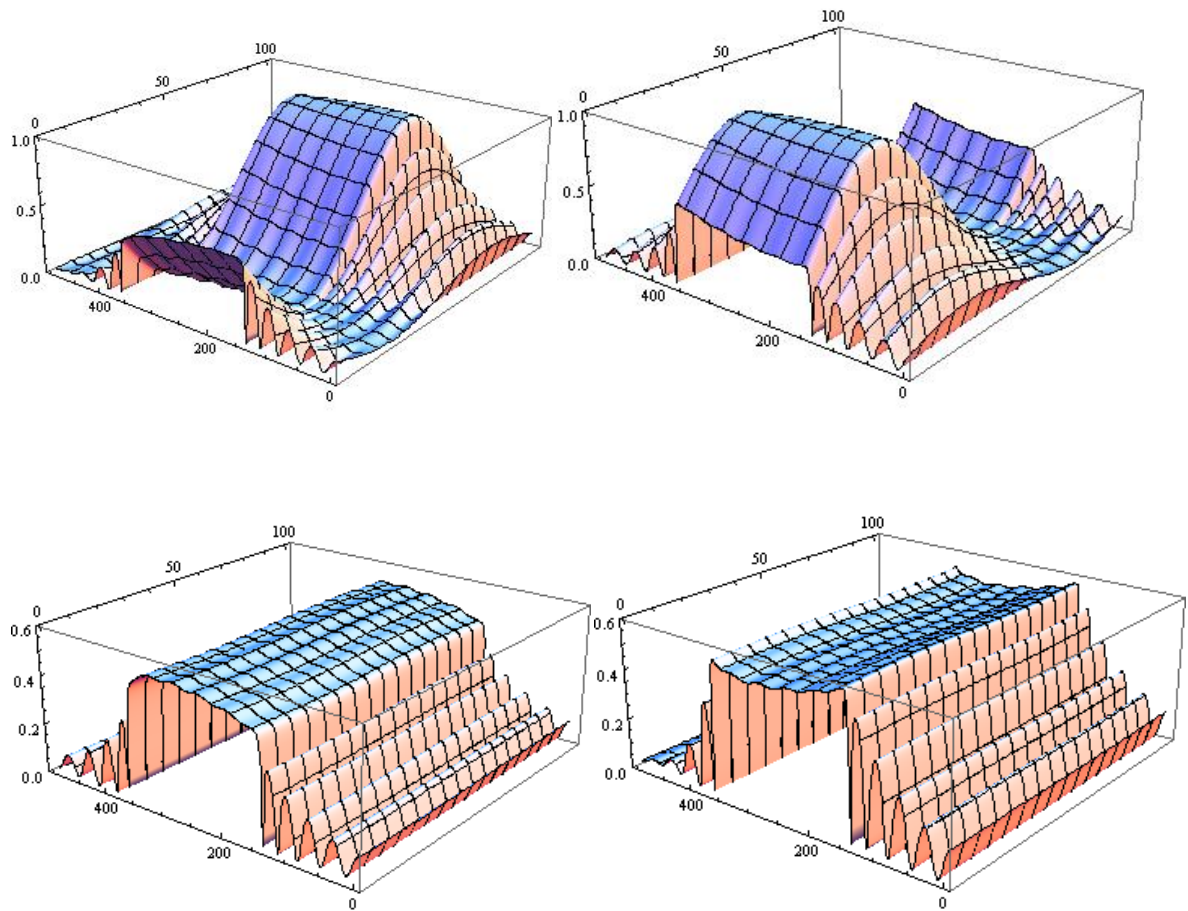

**Fig. 3.** Dependences of the reflection on the wavelength λ and the angle ϕ for the NLC–CLC–NLC system. The incident light has the polarization linear along (a) the *x*-axis and (b) *y*-axis, as well as (c) the right and (d) left circular polarizations. Parameters of the system are the same as in Fig. 2.

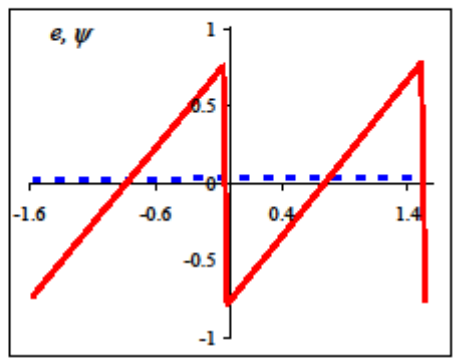

**Fig. 4.** Dependences of (*1*) the azimuth and (*2*) ellipticity of the first EP on the angle ϕ for the wavelength in the center of the PFB (λ = 0.625 μm). Parameters of the system are the same as in Fig. 2.

In Fig. 4 we demonstrate the dependences of the azimuth and ellipticity of one EP on the angle ϕ for a wavelength in the center of PFB. Practically the same dependences are observed for the other EP. Analogous plots are obtained also for other wavelengths in the PFB. Form these data it follows that the NLC–CLC–NLC system may operate also as a system for obtaining the light with an electrically controlled polarization plane rotation. As known, the most of elements of optical systems are commonly polarization-sensitive and change the polarization of light. Each of these elements performs its functions at a given input polarization of light. Therefore it is important (especially in optical communication) to create optical elements for obtaining the signals with a given polarization

We proceed to studying the features of interaction of unpolarized light with the system considered. To describe the interaction of quasi-monochromatic, partially polarized light with optical systems, the formalism of Muller's matrices is commonly used. In this case a solution of the reflection–transmission problem is represented as

$$\mathbf{S}_t = \mathbf{M}_t \mathbf{S}_i, \quad \mathbf{S}_r = \mathbf{M}_r \mathbf{S}_i. \tag{3}$$

Here $\mathbf{S}_i$, $\mathbf{S}_t$, $\mathbf{S}_r$ are the 4×1 Stokes vectors-columns of the incident, transmitted and reflected waves, respectively; $\mathbf{S}_i = I(1, P\cos(2\Phi_i), P\cos(2\Phi_i)\sin(2\Psi_i), P\sin(2\Psi_i))$, $I$ is the total intensity of the incident wave, $\Psi_i$ the azimuth and $\Phi_i$ the ellipticity angle of the polarization ellipse of the completely polarized component in the incident wave, $P$ the degree of polarization of the incident wave, $\mathbf{S}_{r,t} = (S_0^{r,t}, S_1^{r,t}, S_2^{r,t}, S_3^{r,t})$, $S_0^{r,t}$, $S_1^{r,t}$, $S_2^{r,t}$, $S_3^{r,t}$ are the Stokes parameters of the reflected and transmitted waves, correspondingly, $\mathbf{M}_t$ and $\mathbf{M}_r$ are Muller's 4×4 matrices for the transmitted and reflected waves, respectively. From the 2×2 matrices of reflection and transmission, by means of the known rules, one can derive the corresponding Muller's matrices [23]. Intensities of the transmitted and reflected light are determined from the expression $I_{t,r} = S_0^{r,t}$. The azimuths $\theta_{t,r}$ and the angles of ellipticity $\varepsilon_{t,r}$ of the polarization ellipses of the completely polarized components in the transmitted andreflected light are determined from the conditions $\varepsilon_{t,r} = (1/2)arctg\left[S_3^{t,r}/(S_1^{t,r} + S_2^{t,r} + S_3^{t,r})\right]$, $\theta_{t,r} = (1/2)arctg(S_2^{t,r}/S_1^{t,r})$ and the polarization ellipticity from the formula $e_{t,r} = \tan \varepsilon_{t,r}$.

Figure 5 presents the three-dimensional plots of the dependences of the intensity, $tI$ polarization ellipticity $te$ and polarization azimuth $t\theta$ of the transmitted light on the wavelength and angle $\phi$ for the NLC–CLC–NLC system. The incident light is completely unpolarized ($P = 0$). It follows from the plots that the intensity of the transmitted light is approximately equal to a half of the intensity of the incident light, and in the PFB it practically does not depend on the angle $\phi$. The transmitted light is completely polarized, and in the PFB it has a linear polarization, the azimuth of which depends linearly on the angle $\phi$. This means that the NLC–CLC–NLC system can serve as a device for obtaining the linearly polarized light from an unpolarized one, with a possibility to control electrically the polarization azimuth.

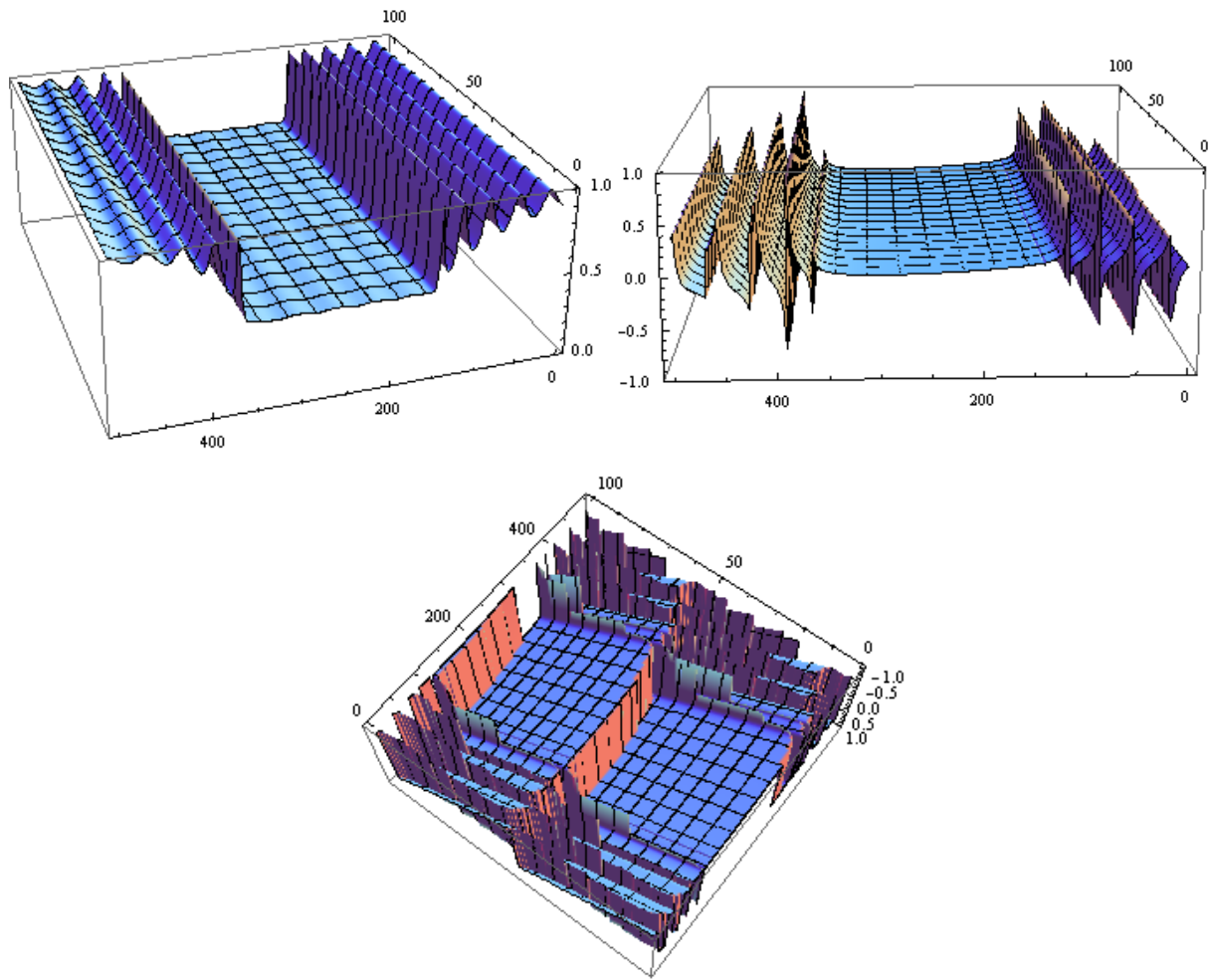

**Fig. 5.** Dependences of (a) the intensity, $I_t$ (b) polarization ellipticity, $e_i$ and (c) polarization azimuth $\theta_t$ of the transmitted light on the wavelength and angle $\phi$. The incident light is completely unpolarized ($P = 0$). Parameters of the system are the same as in Fig. 2